\documentclass{JHEP3}

\input{epsf}
\usepackage{epsfig}
\usepackage{amssymb}
\usepackage{amsfonts}
\usepackage{amsbsy}

\def\vev#1{\langle{#1}\rangle}

\def\BC{\mathbb{C}}
\def\BZ{\mathbb{Z}}
\def\BP{\mathbb{P}}

\def\CO {{\cal O}}

\def\one{{\hbox{ 1\kern-.8mm l}}}

\title{Finiteness of volume of moduli spaces}

\author{Michael R. Douglas$^{1,\&}$ and Zhiqin Lu$^2$\\
$^1$NHETC and Department of Physics and Astronomy\\
Rutgers University\\
Piscataway, NJ 08855--0849, USA\\
$^\&$I.H.E.S., Le Bois-Marie, Bures-sur-Yvette, 91440 France\\
$^2$Department of Mathematics\\
University of California, Irvine 92697\\
{\tt mrd@physics.rutgers.edu, zlu@math.uci.edu}
}

\abstract{
We give a ``physics proof'' of a conjecture made by the first
author at Strings 2005, that the moduli spaces of 
certain conformal field theories
are finite volume in the Zamolodchikov metric, using an RG flow
argument.
}

\begin{document}

\section{A conjecture on moduli spaces}

A rather fundamental question in string/M theory is whether the number
of candidate physical vacua is finite or infinite, and if finite to
estimate its number \cite{JHS60,stat}.  While addressing this question in
all generality is beyond present abilities, it leads to many
relatively well posed subquestions, as the first author
discussed at Strings 2005
\cite{strings}.

One of these questions is whether the volume of any moduli
space of two-dimensional conformal field theories, defined by
integrating the volume form derived from the Zamolodchikov metric, is
finite.  
This type of question was first asked in the physics literature
(as far as we know) by Horne and Moore \cite{HorneMoore}.  They
were motivated by
cosmology, and pointed out that if the total volume were infinite, while
the volume for ``physically viable'' models were finite, there could
be a danger that early cosmological dynamics would never produce the
viable models.  More recently, the question arises because
the volume of moduli space is the leading estimate for the number of
flux vacua in certain string compactifications \cite{DD,DSZ}.  

In the primary examples of sigma models with Ricci flat target
spaces, part or all of the moduli space metric receives no quantum
corrections, and this part 
is the restriction of the standard metric on
the space of metrics to the subspace of Ricci flat metrics.
Thus it is a standard metric studied mathematically, such as the
Weil-Petersson (WP) metric on complex structure moduli space, or a
symmetric space metric.  In many cases, these moduli spaces have already
been shown to be finite volume -- the case of symmetric spaces modulo
arithmetic lattices is well known, and another is the WP metric on the
moduli space of complex structures of a Calabi-Yau threefold
\cite{Todorov1,LuSun1, LuSun}. For more on the Weil-Petersson geometry, we 
refer to 
\cite{Todorov, LuSun1}.

We only know one counterexample to the claim, namely the $c=1$ bosonic
and the $(1,1)$ $\hat c=1$ SCFT's, {\it i.e.} sigma models with target
space $S^1$.  Of course, if we consider volumes of submanifolds of a moduli
space as well, the existence of infinite length paths in moduli space
is well known.  The basic example of how this is compatible with
finite volume is $T^2$ target space, as discussed in \cite{HorneMoore}.

It is not obvious to us whether the case of target space $S^1$ is an
isolated exception, or whether it signals a general class of examples.
Even if there were other examples, if they were all associated with
the large volume limit, they would not be a problem for the finiteness
question posed in \cite{stat,strings}, as there we require finiteness
of the number of vacua with a lower bound on the KK scale.

A related and somewhat simpler conjecture is finiteness of the volume
of the part of CFT moduli space in which there is a lower bound on
non-zero operator dimensions, which we will call the ``gap.''%
\footnote{We understand that Cumrun Vafa has also considered this
variation, in a paper which appeared as this one was finished.}
Our original motivation for considering this version is that
a similar condition, namely an upper bound on the diameter, is often
made when discussing related questions in Riemannian geometry.  In the
Strings talk, we explained this in the context of Cheeger's theorem,
and we discuss this further in \cite{AchDoug}.  Another motivation for
this form of the conjecture is Kontsevich's work on CFT
\cite{Kont-discuss}.  He has conjectured that the space of CFT's with
fixed $c$ and a lower bound on the gap is precompact, meaning that
there is a distance function on the space such that any infinite
sequence has a Cauchy subsequence.  While this is a weaker,
topological claim, it is far more general in the sense that one is not
limited to talking about moduli spaces; one can talk about any
sequence of CFT's.

In the following, we outline a physical argument for finiteness given
a gap, inspired by work of the second author, and which we are
pursuing in joint work to appear, along with discussions of the
related mathematics and an attempt to compute the actual volume of the
moduli space of quintic hypersurfaces.

\section{An RG approach}

At present there is no general definition of CFT which is concrete
enough to address the conjecture in general, but a very useful
definition for certain cases, such as Calabi-Yau target space, is the
gauged linear sigma model (GLSM) \cite{Witten}.  This produces a CFT
as the IR limit of an RG flow from a non-trivial but weakly coupled UV
theory, and thus the UV limit of the Zamolodchikov metric is
computable in perturbation theory.  Thus, we adopt a two-step
approach, of first showing that the UV metric is finite volume, and
then arguing that the RG flow will preserve this finiteness.

While the argument is general,
for concreteness, let us discuss the case of the quintic hypersurface
in $\BC\BP^4$.  The GLSM is now a $(2,2)$ sigma model with a
$U(1)$ gauge symmetry and six chiral superfields, five of $U(1)$ charge
$Q=+1$ denoted $Z^i$ for $1\le i\le 5$, and a sixth of $U(1)$ charge
$Q=-5$ denoted $P$.  Finally, there is a generic superpotential of
degree $5$ in $Z$ and linear in $P$.

The bare couplings in the model include $126$ coefficients in the
superpotential, the gauge coupling (a relevant operator), and a
Fayet-Iliopoulos term for the $U(1)$ which we denote $\zeta$,
complexified by a ``theta angle'' $\theta$
controlling the operator $\int F$.  In addition, one can vary the
kinetic term; the only operators here which are not irrelevant are
the matrix entering in the quadratic term $g_{i\bar j} Z^i \bar Z^{\bar j}$
and the $P\bar P$ term.  These are redundant with the overall scale
of the superpotential coefficients but we will keep them for now.

The basic structure of the RG flow is simply a flow down to the
nonlinear sigma model whose target space is the moduli space of
``vacua'' (constant solutions of the equations of motion).  This can
be found in two steps.  First, solving the D-term conditions and
quotienting by $U(1)$, one finds two results depending on the sign of
$\zeta$.  For $\zeta>0$, the quotient is the noncompact Calabi-Yau
fivefold $\CO_{\BC\BP^4}(-5)$, while for $\zeta<0$ it is the orbifold
$\BC^5/\BZ^5$.  The first is just the blowup of the second at zero and
is the only crepant resolution of this orbifold singularity, so the
two possibilities are closely related in algebraic geometric terms.
Second, one considers the superpotential constraint $W'=0$ in the two
cases.  In the case of $\CO_{\BC\BP^4}(-5)$, the F-term constraint
$\partial W/\partial P=0$ defines a quintic hypersurface, while the
others force $P=0$ so the hypersurface sits in $\BC\BP^4$.  
In the case of $\BC^5/\BZ^5$, one has $P\ne 0$,
so the conditions $\partial W/\partial Z^i=0$ are nontrivial and
effectively produce a Landau-Ginzburg (LG) model, whose $\BZ^5$
quotient is the Gepner model construction.  One can check that the two
regimes of $\zeta$ are continuously connected if $\theta\ne 0$, and
this was Witten's original proof of the CY-LG equivalence.

A slightly more detailed picture of the RG flow can be obtained by
noting that the order in which these two steps (D and F term)
are implemented depends on the ratio of the energy scales set by the
gauge coupling (which controls the D-term potential) and the
overall coefficient superpotential.  Both are relevant operators
and thus one expects their couplings
(both of mass dimension $2$)
to set these scales directly.

Far above both scales, perturbation theory is good, and we can
compute a good approximation to the Zamolodchikov metric in free
field theory.  We will concentrate on the metric on complex structure
moduli space; the rest can be computed as well and is easily seen
to decouple from the complex structure metric in leading order.

A basis for the variations of complex structure are the $126$
operators
$$
O_I \equiv O_{i_1i_2i_3i_4i_5} = P Z^{i_1} Z^{i_2}Z^{i_3}Z^{i_4}Z^{i_5} .
$$
whose couplings $t^I$ parameterize $V\equiv\BC^{126}$.  In free field
theory,
$$
\vev{Z^i(1)\bar Z^{\bar j}(0)} = g^{i\bar j} G(1,0); \qquad
\vev{P(1)\bar P(0)}=G(1,0) ,
$$
where $g^{i\bar j}$ is the inverse of the metric appearing in the
bare $Z\bar Z$ kinetic term, and
$G(z_1,z_2)$ is the two-point function for a free boson.  Multiplying
these to get the two-point function and Zamolodchikov metric
$$
G_{I\bar J} = \vev{O_I(1) \bar O_{\bar J}(0)} ,
$$
one finds that this metric is diagonal and non-degenerate, with
coefficients given by simple combinatoric factors which will not
concern us below.

The true complex structure moduli space is the $101$ complex
dimensional quotient of $V\equiv\BC^{126}$ by a $GL(5)$ action induced
by the linear action on the vector $Z^i$.  This $GL(5)$ symmetry is
broken to $U(5)$ by any explicit choice of the bare metric $g_{i\bar
j}$ and thus its restoration in the IR is a result of the RG flow,
which if non-singular will lose the information of the choice of bare
coupling $g_{i\bar j}$.  Thus, to relate the volume of the moduli
space between UV and IR, we must choose a $101$-dimensional slice
through the space of bare couplngs $V$ which intersects each orbit
once.  The metric on this slice will then be obtained by restricting
the original Zamolodchikov metric on $V$.  Dependence on the choice of
slice should of course drop out in the IR.

Such a slice can be naturally obtained by symplectic reduction, {\it i.e.}
imposing constraint equations 
$$
1 = G_{I\bar J} t^I \bar t^{\bar J}
$$
and quotienting by the $U(5)$ which preserves $G_{I\bar J}$.  This can
be done in two steps, first quotienting by the $GL(1)$ acting as an
overall rescaling of the fields, and then by $SL(5)$.  The result of the
first quotient is the compact moduli space $\BC\BP^{125}$
with a manifestly nonsingular metric, similar to the Fubini-Study metric,
and thus its volume is manifestly finite.  The second quotient can introduce
singularities in general but nevertheless the volume remains finite.

Before moving on, we recall that not all points in $\BC\BP^{125}//SL(5)$
correspond to non-singular Calabi-Yau manifolds.  For this to be the
case, the defining equation $f=0$ must be non-singular, meaning that
$\partial f/\partial Z^i\ne 0$ at every point such that $f=0$.  This
condition defines an open subset of the quotient space whose
complement is a codimension one subvariety called
the discriminant locus.  Now the actual Weil-Petersson metric on complex
structure moduli space is typically singular on the discriminant locus,
and thus the claim that the moduli space volume in this metric is finite
is rather subtle as one needs to control this divergence.  However, the
UV approximation we have computed so far is nonsingular at these points.
This is no surprise as physically the singularities are related to
properties of the Calabi-Yau metrics in these regions of moduli space
(for example, the formation of long throat regions) which are not true 
of the UV metric.  It does mean that we should not trust an argument
for finiteness which does not address this subtlety in some way.

\subsection{The RG flow}

We hvae just seen that the volume of complex moduli space in the UV is
finite.  We now need to argue that this property is preserved under RG
flow.

The basic physical intuition for this is that the metric will only
flow by a finite amount in any finite amount of RG ``time'' (ratio of
energy scales), and non-trivial RG evolution takes place only over a
finite time.  In principle, this would be made precise by using the RG
flow equations for the couplings, to derive an RG flow equation for
the Zamolodchikov metric.  

We first note that standard nonrenormalization theorems imply that the
complex structure moduli $t^I$ themselves do not flow.
The flow equation for the Zamolodchikov metric takes the form
$$
\mu\frac{\partial}{\partial\mu} G_{I\bar J}
 = \gamma_{I\bar J}(t,\rho) \sim \gamma_I(t,\rho)+\gamma_{\bar J}(t,\rho).
$$ 
Note that this is not literally a beta function as $G_{I\bar J}$ is
not a sigma model coupling but a two-point function.  However its
dependence under coordinate rescaling amounts to a flow, which in some
approximation is given by the sum of the anomalous dimension
coefficients $\gamma_I$ for $O_I$ and $\gamma_{\bar J}$ for $\bar
O_{\bar J}$.  In general, this flow will depend on both the point in
complex structure moduli space $t$, and on the K\"ahler moduli $\rho$.
Iit can produce a fairly arbitrary metric after a finite flow;
let us denote this as
$$
G_{I\bar J}(\mu) \equiv \vev{O_I(\mu^{-1}) \bar O_{\bar J}(0)} .
$$
Indeed,
the simple metric we derived in the UV need have little resemblance to
the actual Weil-Petersson metric we expect to recover in the IR.

However, given that the anomalous dimensions are finite, the change
in any given metric coefficient induced by a finite flow will be finite,
$$
\frac{G_{I\bar J}(\mu_1)}{G_{I\bar J}(\mu_2)}
 \sim \left(\frac{\mu_1}{\mu_2}\right)^\gamma
$$
where $\gamma=0$ for a truly marginal operator and might be expected
to be typically (though not always) small.

While the total RG time is infinite, one can see that most of the flow
takes place around the energy scales $g$ set by the gauge coupling and
$|W|$ set by the superpotential, and is small in the extreme UV and IR.
In the UV, this follows from standard results on the behavior of
perturbation theory, while in the IR, it follows from the assumption that,
at a given point in moduli space $(t,\rho)$, the flow to the IR fixed
point comes in along a leading irrelevant direction controlled by an
operator with dimension $2+\epsilon$ with some $\epsilon>0$.  In this
case, the estimate for the total flow in any coupling or correlation
function induced by a flow from a scale $\mu_0$ at which this assumption
is valid, to the IR, is
$$
G_{I\bar J}({\rm IR}) \sim G_{I\bar J}(\mu_0) + O(\mu_0^\epsilon).
$$
Thus this is again expected to be finite.

This argument can be applied pointwise in the moduli space and thus 
leads to finiteness, if its assumptions are satisfied.  Conversely,
it suggests that any possible divergence in the Zamolodchikov metric
as we approach some limiting point $p$ in moduli space,
and thus a divergence in the volume form, can be traced back to
the fact that the RG time required for the flow diverges as we approach
$p$.  The other possibility, that beta function coefficients themselves
diverge, seems unphysical.

Such a divergence of the RG flow time could appear for at least two
reasons.  First, it could be that at $p$, the RG flow actually does
not reach a $\hat c=3$ SCFT, but instead stops at an intermediate
fixed point with $3<\hat c<6$.  In this case, as we approach $p$,
the RG flows will get ``hung up'' near the intermediate fixed point,
for a time which grows as we approach $p$ as a power controlled by the
dimension of the leading relevant operator coming out of $p$. 
We do not know of evidence that this happens in the case at hand,
but it is an interesting possibility.

Second, it could be that the assumption that there is a gap
$\epsilon>0$ to the leading irrelevant operator fails, and the flow
takes infinite time in the IR.  Clearly this assumption can fail; it
is not at all obvious even in the large volume limit (there is no
shortage of candidate operators there), and presumably must fail near
a conifold point, and
other limits in which metric coefficients can diverge.  Indeed, 
in the degeneration limit of the conifold, 
the gap goes to zero \cite{Kutasov}, as follows from a
candidate $\hat c=3$
SCFT description of the noncompact CY containing the conifold region
\cite{Giveon:1999px}.  Thus this mechanism can produce
the known logarithmic divergence in the moduli space metric.

Thus, up to the possible existence of intermediate fixed points, we
have ``physically proven'' a weak form of the finiteness conjecture,
that the volume of moduli space in regions with a gap is finite.

Since already the conifold CFT does not have a gap, and it presumably
can be part of a physically realistic compactification (one is not
sending the KK scale to zero but rather a scale associated with the
throat region, which could decouple from the physical sector), this is
rather weaker than the conjecture in \cite{strings}, but is clearly a
step in this direction, as it shows that the ``bulk'' of moduli space
has finite volume, and relates moduli space properties to the RG in an
interesting way.  We believe that incorporating more of the
mathematics which enters the proof of the finiteness of the total
moduli space volume will lead to a better physical argument, and
are pursuing this.

We conclude by mentioning a much stronger claim which is 
suggested by the work of the second author and which we are also
exploring in our ongoing work.  It is that the RG flow preserves the
K\"ahler class of the moduli space metric, in the sense that the
difference of the two K\"ahler forms $\omega_{IR}-\omega_{UV}$ is
exact and sufficiently weakly singular that the volume does not
contain boundary terms.  Again, there are simple arguments for the
exactness, which we will give in \cite{toappear}; the subtle part of
this claim is the weakness of the singularity, which allows computing
the total volume of the moduli space by integrating the very simple
volume form $\omega_{IR}^n/n!$.  This point may be within mathematical
reach and would allow us to estimate this total volume, completing the
estimate of the asymptotic number of attractor points begun in
\cite{DD} for a high dimensional case.  Another important
generalization would be to discuss the curvature form of \cite{AD}.

We thank D. Friedan, A. Konechny, M. Kontsevich, D. Kutasov, G. Moore,
T. Pantev and C. Vafa, for helpful discussions.

The research of M.R.D. is supported by DOE grant DE-FG02-96ER40959,
while that of Z.L. is supported by NSF grant DMS 0347033..

\end{document}